\documentclass[letterpaper]{article}
\usepackage{aaai25}
\usepackage{times}
\usepackage{helvet}
\usepackage{courier}
\usepackage[hyphens]{url}
\usepackage{graphicx}
\usepackage{natbib}
\usepackage{caption}
\usepackage{lipsum}

\usepackage[linesnumbered,lined,ruled,vlined]{algorithm2e}
\usepackage{algpseudocode}

\usepackage{amsmath,amssymb,amsfonts}
\usepackage{mathtools}
\usepackage{booktabs}
\usepackage{tabularx}
\usepackage{enumitem}
\usepackage{graphicx}
\usepackage{subcaption}
\usepackage[utf8]{inputenc}
\usepackage{multirow}
\usepackage{array}

\title{Hierarchical Deep Reinforcement Learning for Adaptive Resource Management in Integrated Terrestrial and Non-Terrestrial Networks}
\author {
    Muhammad Ahmed Mohsin\textsuperscript{\rm 1},
    Hassan Rizwan\textsuperscript{\rm 2},
    Muhammad Umer\textsuperscript{\rm 3},
    Sagnik Bhattacharya\textsuperscript{\rm 1},\\
    Ahsan Bilal\textsuperscript{\rm 4},
    John M. Cioffi\textsuperscript{\rm 1}
}

\affiliations {
\textsuperscript{\rm 1}Stanford University, 450 Jane Stanford Way, Stanford, CA 94305-2004, USA \\
\textsuperscript{\rm 2}University of California, Riverside, 900 University Ave, Riverside, CA 92521, USA \\
\textsuperscript{\rm 3}National University of Sciences and Technology, Sector H-12, Islamabad, Pakistan \\
\textsuperscript{\rm 4}University of Oklahoma, 660 Parrington Oval, Norman, OK 73019, USA \\
muahmed@stanford.edu,
hrizwan@email.com,
mumer.bee20seecs@seecs.edu.pk,
sagnikb@stanford.edu,\\
ahsan.bilal-1@ou.edu,
cioffi@stanford.edu
}

\begin{document}
\maketitle

\begin{abstract}
  Efficient spectrum allocation has become crucial as the surge in wireless-connected devices demands seamless support for more users and applications, a trend expected to grow with 6G. Innovations in satellite technologies such as SpaceX's Starlink have enabled non-terrestrial networks (NTNs) to work alongside terrestrial networks (TNs) and allocate spectrum based on regional demands. Existing spectrum sharing approaches in TNs use machine learning for interference minimization through power allocation and spectrum sensing, but the unique characteristics of NTNs like varying orbital dynamics and coverage patterns require more sophisticated coordination mechanisms. The proposed work uses a hierarchical deep reinforcement learning (HDRL) approach for efficient spectrum allocation across TN-NTN networks. DRL agents are present at each TN-NTN hierarchy that dynamically learn and allocate spectrum based on regional trends. This framework is 50x faster than the exhaustive search algorithm while achieving 95\% of optimum spectral efficiency. Moreover, it is 3.75x faster than multi-agent DRL, which is commonly used for spectrum sharing, and has a 12\% higher overall average throughput.
\end{abstract}

\section{Introduction}

The increasingly complex ecosystem of wireless networks---from cellular systems to satellite constellations, Internet of things (IoT) devices to vehicular networks---operating within a constrained spectrum band necessitates efficient spectrum sharing. Spectrum sharing refers to the dynamic allocation and reuse of radio frequency bands among multiple users, particularly in complex interference channel (IC) scenarios where multiple transmitters and receivers operate simultaneously. Efficient spectrum sharing reduces co-channel interference and thereby maintains a higher signal-to-interference-plus-noise ratio (SINR) and preserves reliable network throughput for communication. Technological breakthroughs in satellite constellation deployments, led by innovators such as SpaceX's Starlink, Amazon's Project Kuiper, and OneWeb, transform wireless communication landscapes. These non-terrestrial networks (NTNs) now seamlessly coexist with terrestrial networks (TNs), establishing intricate, multi-tiered network architectures that operate across diverse altitudinal ranges. Globally, the number of operational satellites is expected to increase from 13,000 to 33,000 by 2030, enabling 20-30\% of mobile users to shift to satellite internet services~\cite{ceoworld2024}. This market is projected to grow from \$9 billion in 2023 to \$37 billion by 2034~\cite{precedenceresearch2024}. Therefore, developing efficient spectrum sharing mechanisms for integrated TN-NTN has become crucial to ensure resource optimization and seamless coexistence between these diverse network architectures.

Deep reinforcement learning (DRL) emerges as a particularly compelling solution for optimal spectrum sharing due to its ability to adapt to complex wireless environments and learn optimal policies through continuous interaction~\cite{10024896}. Unlike traditional machine learning approaches that rely on static training datasets and struggle to generalize beyond their training distributions, DRL agents can dynamically adjust their spectrum allocation strategies based on real-time network conditions, interference patterns, and quality of service (QoS)~\cite{zhang2024drl}.

Existing spectrum sharing strategies often oversimplify spectrum allocation as they fail to account for the nested hierarchy present in modern network architectures~\cite{patil2023comprehensive}. These approaches demonstrate efficacy in controlled environments but, their centralized nature introduces a significant overhead in execution time and leads to suboptimal outcomes due to computational bottlenecks and delays in gathering state information~\cite{zhang2023review}. In large-scale deployments with heterogeneous architectures, spectrum sharing faces severe scalability limitations as the complexity of interference management and resource allocation in ICs grows exponentially with expanding network size and user density. Traditional solutions have primarily focused on TNs, employing conventional techniques like power control~\cite{9681985}, interference management~\cite{9802083}, and spectrum sensing~\cite{9044839} for efficient spectrum usage. However, these approaches fail to address the unique challenges of TN-NTN which require coordinated decision-making across multiple network tiers.




This paper proposes a novel hierarchical deep reinforcement learning (HDRL)-based spectrum allocation scheme for integrated TN-NTN networks. The network architecture is hierarchically decomposed into three distinct sub-networks: satellite, high altitude platforms (HAPs), and a combined layer of unmanned aerial vehicle (UAVs) and terrestrial base stations (TBSs). Each sub-network's agent operates on a different temporal scale with interconnected policies, where higher-level agents guide the behavior of lower-level agents through metacontrol signals. The hierarchical structure ensures that each subsequent network layer operates within the spectrum constraints imposed by its preceding layer, creating a cascaded decision-making framework that maximizes spectrum utilization across the entire network.

\begin{itemize}

  \item The proposed framework was benchmarked against different algorithms, including exhaustive search, random access, single-agent DRL (SADRL), and multi-agent DRL (MADRL), across three network hierarchies. The framework achieved 95\% of the spectral efficiency of exhaustive search while being 50x faster. It also demonstrated 3.75x faster execution than MADRL and yielded 10-18\% performance improvements over SADRL and random access methods across all scenarios.

  \item In a single 500-step episode, the framework achieved 5\%, 11\%, and 25\% higher average throughput compared to MADRL, SADRL, and random access, respectively. The framework maintained superior stability with minimal throughput fluctuations across all steps relative to both MADRL and SADRL.

  \item The framework's learning progression and convergence behavior were evaluated across different network hierarchies to assess adaptability. Training results over 1000 episodes showed consistent learning progress, with the space-air-ground (SAG) network achieving the highest cumulative reward, followed by the air-ground and UAV-aided networks.

\end{itemize}

\section{Related Work}

\subsection{Spectrum Sharing}

Several previous works focus on interference management for spectrum sharing in satellite-terrestrial networks:~\cite{9686170} introduces reverse spectrum pairing for TN-NTN systems, while~\cite{10301688} optimizes TN-NTN grouping with earth-fixed satellite beamforming. To accommodate different network architectures,~\cite{9061002} develops a cognitive control system for air ground integrated networks (AGIN) spectrum sharing, whereas,~\cite{9163369} and~\cite{8633412} focus on satellite spectrum sharing frameworks---the former for geostationary earth satellites (GEO) and low-orbit earth satellites (LEO) networks using overlay/underlay modes, and the latter for satellite-terrestrial mmWave networks using protection areas. These works allow spectrum sharing via interference mitigation through spectrum sensing and power control mechanisms, however, they overlook the challenge of spectrum allocation across integrated TN-NTN networks where spectrum resources need to be dynamically distributed based on regional demand patterns.

\subsection{DRL for Spectrum Management}

There are several works exploring DRL approaches for spectrum management:~\cite{9336707} and~\cite{9557877} utilize deep Q-network (DQN) variants for dynamic spectrum access (DSA) coordination, with the latter implementing dueling DQNs and prioritized experience replay for balanced primary user and secondary user performance.~\cite{9541168} introduces a scalable single-agent reinforcement learning (RL) method for joint routing and spectrum optimization in wireless ad-hoc networks. For distributed approaches,~\cite{9759480} employs cooperative multi-agent RL with recurrent DQNs for DSA, while~\cite{s22041630} proposes multi-agent DQL for HAPs power control with interference management.~\cite{10001062} addresses urban air mobility using DRL-based spectrum allocation between aerial and terrestrial users.

While these approaches demonstrate the effectiveness of utilizing DRL algorithms for spectrum sharing, they primarily focus on TNs and single-layer optimization, neglecting the spectrum management required for NTNs in conjunction with TNs, and the need for hierarchical decision-making across multiple network layers.

\section{HDRL Based Intelligent Spectrum Allocation}

\subsection{System Overview}

The considered system comprises a LEO satellite, HAPs, UAVs, TBSs, and users. The LEO satellite covers designated geographical regions using fixed multi-beam technology, where each beam cell serves as a dedicated coverage area for a HAP. Each HAP acts as a regional hub, relaying data and control signals between the satellite and lower-tier network nodes. Within each HAP’s coverage area, multiple TBSs and UAVs are deployed. TBSs provide fixed, high-capacity connectivity for ground-based users, while UAVs act as aerial base stations, offering flexible, on-demand coverage. Users dynamically associate with either a TBS or UAV based on factors such as signal strength, network load, and QoS. A satellite gateway facilitates communication between the TN and the LEO satellite while a central control and compute unit manages network operations, acting as the coordinator for spectrum allocation and interference management. The LEO satellite allocates portions of its spectrum \( A_{\text{satellite}} \) to each beam cell. Each beam's spectrum is then further divided by the HAPs into subbands \( A_{\text{HAP}} \) and is shared by both UAVs and TBSs in that coverage region.



\subsection{Problem Formulation}

\begin{figure*}[t!]
  \centering
  \includegraphics[width=0.995\textwidth]{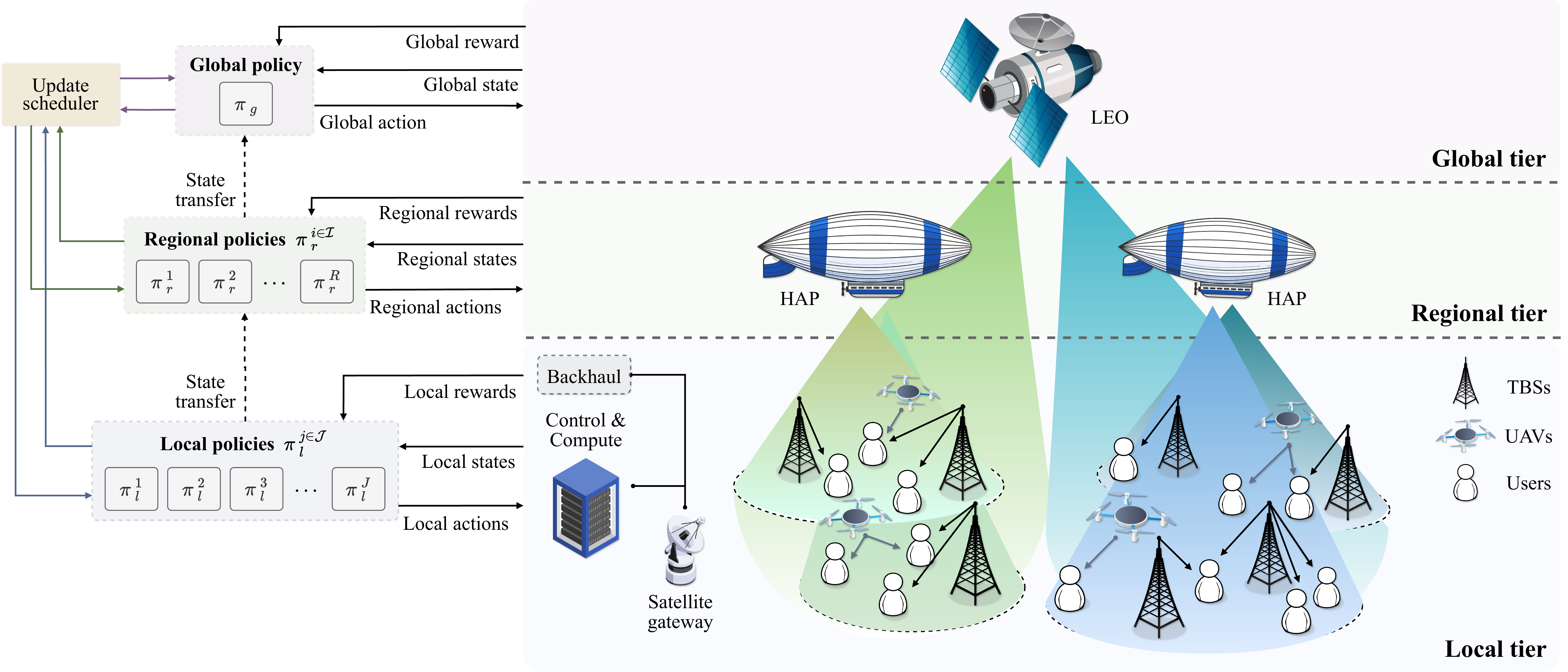}
  \caption{Proposed framework environment for HDRL-based dynamic spectrum sharing in integrated TN-NTN Networks.}
  \label{fig:enter-label}
\end{figure*}

The hierarchical spectrum sharing framework is modeled as a Markov decision process (MDP), captures the multi-tiered decision-making process essential for efficient spectrum allocation in a TN-NTN system. The MDP consists of a state space \( S \), an action space \( A \), a state transition function \( s' = f(s, a) \), and a reward function \( r(s, a) \). DRL agent interacts with the environment over a sequence of states \( s_1, s_2, \dots, s_t \), actions \( a_1, a_2, \dots, a_t \), and rewards \( r_1, r_2, \dots, r_t \), where \(s_t, a_t, r_t\) are state, action and reward at time \(t\) and total time steps in the episode are \(T\). The primary objective is to optimize network performance across three hierarchical decision levels---satellite, HAP, and UAV---each responsible for spectrum management within its respective operational scope.

\subsubsection{Global Policy.}
At the top level, the \textit{global policy} \( \pi_g \) is managed by the satellite agent, which oversees the entire network spectrum allocation. This agent’s goal is to allocate spectrum resources effectively across multiple beam cells, ensuring fair distribution and accommodating varying user demands and channel conditions. The global network state, denoted as
\[
  S_g = \{A_{\text{spec}}, D_{\text{beam}}, G_{\text{avg}}\},
\]
includes aggregated information such as the total available spectrum \( A_{\text{spec}} \), the distribution of beams \( D_{\text{beam}} \) across geographical regions, and the average channel gain \( G_{\text{avg}} \) across these regions. Given this state \( S_g \), the satellite agent determines a spectrum allocation matrix \( \mathbf{A}_g \in [0,1]^{B \times N} \), where each element \( a_{b,n} \) represents the allocation of subband \( n \) to beam \( b \)
\begin{equation}
  a_{b,n} = \begin{cases}
    1 & \text{if subband } n \text{ is allocated to beam } b \\
    0 & \text{otherwise}
  \end{cases}
\end{equation}
This allocation must satisfy
\begin{equation}
  \sum_{b=1}^B a_{b,n} \leq 1, \forall n \in \{1,\ldots,N\}.
\end{equation}

The selected action \( a_g = \mathbf{A}_g \) is constrained by the policy \( \pi_g(S_g) \), which optimizes the high-level allocation to maximize network performance metrics like spectral efficiency \( \eta \), system fairness \( F \), and average throughput \( R_{\text{avg}} \). This global level allocation subsequently passes down to the regional tier as a constraint on available resources, ensuring that the regional and local policies can operate within these allocations.

\subsubsection{Regional Policies.}

At the intermediate level, each HAP \( i \) manages a \textit{regional policy} \( \pi_r^i \), responsible for spectrum allocation within its designated coverage area. Given the global allocation constraints from \( \pi_g \), each HAP agent operates on a regional network state \( S_r^i \), which incorporates detailed information relevant to the local context. This state is defined as
\[
  S_r^i = \{A_{\text{spec}}, D_{\text{region}}, G_{\text{avg}}\},
\]
where \( A_{\text{spec}} \) is the spectrum allocated by the global policy to the HAP's region, \( D_{\text{region}} \) represents the spatial distribution of users within the HAP's coverage, and \( G_{\text{avg}} \) indicates average channel conditions in the region.

\begin{figure*}[t!]
  \centering
  \includegraphics[width=0.675\linewidth]{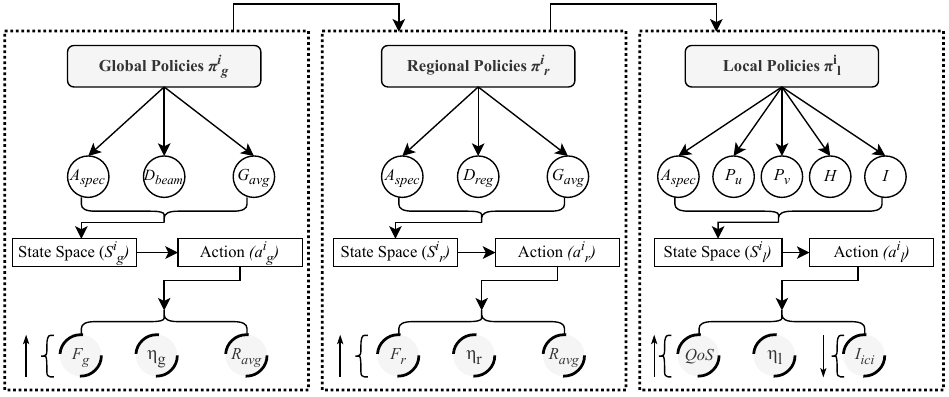}
  \caption{Flow diagram for the proposed hierarchical reinforcement learning for spectrum resource management.}
  \label{fig:flowchart}
\end{figure*}

At this level, each HAP determines a regional spectrum allocation matrix \( \mathbf{A}_r^i \in [0,1]^{M \times N} \), where \( M \) is the number of subordinate nodes (UAVs and TBSs) in region \( i \). The allocation elements are
\[
  a_{m,n}^i = \begin{cases}
    1 & \text{if subband } n \text{ is allocated to node } m \text{ in region } i, \\
    0 & \text{otherwise},
  \end{cases}
\]
which are subject to the constraints
\begin{align}
  \sum_{m=1}^M a_{m,n}^i & \leq 1, \quad \forall n \in \{1,\ldots,N\}, \\
  a_{m,n}^i              & \leq a_{b(i),n}, \quad \forall m,n,
\end{align}
where \( b(i) \) denotes the beam containing region \( i \).

\subsubsection{Local Policies.}

At the lowest level, individual UAVs and TBSs operate under \textit{local policies} \( \pi_l^j \) for each node \( j \), making real-time decisions on spectrum access and power allocation for their associated users. The local network state \( S_l^j \) for each UAV or TBS agent \( j \) is represented as
\[
  S_l^j = \{A_{\text{spec}}, P_u, P_v, H, I\},
\]
where \( P_u \) and \( P_v \) indicate the positions of the user equipment (UE) and UAV, respectively, \( H \) represents the channel gains, and \( I \) denotes interference levels in the immediate vicinity.
Based on \( S_l^j \), each local policy \( \pi_l^j \) selects an action \( a_l^j \), which is represented as
\[
  a_l^j = \{\beta_j, \alpha_j, \Delta p_j\},
\]
where the spectrum access vector \( \beta_j \in [0,1]^N \), the power allocation vector \( \alpha_j \in [0,1]^K \), and the movement vector \( \Delta p_j \in [-\Delta p_{\text{max}}, \Delta p_{\text{max}}]^2 \). The spectrum access vector \( \beta_j \) indicates which spectrum channels to utilize, allowing the agent to decide the specific channels that are most efficient under current network conditions. The power allocation vector \( \alpha_j \) specifies the power levels assigned to each selected channel, ensuring that power resources are optimally distributed to minimize interference while meeting user demands. Lastly, for UAVs, the movement vector \( \Delta p_j \) controls positional adjustments within the UAV’s operational range, allowing it to enhance local coverage dynamically by repositioning itself in response to user distribution and channel quality variations. Penalty $P_{UAV}$ is imposed if the UAVs are outside their operational range
\begin{equation}
  P_{UAV} = \frac{\sum_{u=1}^U \mathbb{1}(u \notin \mathcal{R})}{U},
\end{equation}
where $\mathbb{1}(u \notin \mathcal{R})$ is the indicator function for UAVs outside their designated region, and $U$ is the total number of UAVs per region. This local-level optimization ensures that real-time adjustments to spectrum access and power control are responsive to changes in user demand, channel conditions, and interference levels.

At the regional level, several metrics are evaluated to determine the performance of each region $j \in J$, where $J$ represents the set of all local regions. We utilized appropriate channel distributions for all channel links. Each $u$ user's SINR is defined as
\begin{equation}
  \label{SINR}
  \gamma_{ju} = \frac{H_{ju} \alpha_{ju} P}{I_{ju} + N_0},
\end{equation}
where \( H_{ju} \) represents the channel gain, \( \alpha_{ju} \) is the power allocation, \( P \) is the transmission power, \( I_{ju} \) denotes interference, and \( N_0 \) is the noise power. Using~\eqref{SINR}, the data rate for each user is
\begin{equation}
  R_{ju} = \frac{W}{N} \log_2 (1 + \gamma_{ju}),
\end{equation}
where \( W \) is the total bandwidth available and \( N \) is the number of spectrum channels. We compute spectral efficiency in each region as
\begin{equation}
  \eta_j = \frac{\sum R_{ju}}{W},
\end{equation}
where \( \eta_j \) represents the spectral efficiency. To measure fairness, we calculate the Jain's fairness index for each region as
\begin{equation}
  F_j = \frac{\left( \sum R_{ju} \right)^2}{K \sum R_{ju}^2},
\end{equation}
where \( K \) is the number of users in region \( j \). Finally, to monitor QoS adherence, we calculate the QoS violation for each region as
\begin{equation}
  V_j = \max(0, R_{\text{min}} - \min(R_{ju})),
\end{equation}
where \( R_{\text{min}} \) is the minimum required rate for QoS compliance. These metrics collectively provide a detailed assessment of each region's performance and are instrumental in formulating the reward function.

\subsection{Optimization Objective}

The hierarchical structure of the MDP maximizes the cumulative reward over all time steps $t$ formulated as
\begin{equation}
  \max_{\pi_g, \pi_r, \pi_l} \mathbb{E} \left[ \sum_{t=1}^T \left( w_1 R_{\text{avg}} + w_2 \eta + w_3 F + w_4 P_{\text{UAV}}\right) \right],
\end{equation}
where \( w_1 \), \( w_2 \), \( w_3 \), and \( w_4 \) are weights for data rate, spectral efficiency, fairness, and, UAV penalty respectively. This reward structure incentivizes each policy level to contribute towards optimal network performance, balancing local needs and global objectives in a dynamically adaptive manner. Through this hierarchical MDP structure, the HDRL framework facilitates efficient spectrum sharing by decomposing the overall optimization problem into manageable sub-problems, each tailored to the operational scope and constraints of the respective agent.

  {\setlength{\algomargin}{1.25em}
    \begin{algorithm}[t!]
      \DontPrintSemicolon
      \caption{HDRL for Spectrum Sharing}\label{alg:hdrl}

      Initialize neural networks for policies $\pi_s$, $\pi_h$, $\pi_l$\;
      Initialize replay buffers $D_s$, $D_h$, $D_l$\;
      \For{episode $= 1$ to $M$}{
        Initialize parameters $H$, $I$, $P_u$, $P_v$, $\alpha$\;

        \For{step $= 1$ to $S$}{

          \If{$step~\%~\delta_s = 0$}{
            Observe (beam) $s_s = \{A_{\text{spec}}, D_{\text{beam}}, G_{\text{avg}}\}$\;
            Take action $a_s = \pi_s(s_s)$ for $A_{\text{beam spec}}$\;
          }

          \If{$step~\%~\delta_h = 0$}{
            \For{each HAP $i = 1$ to $B \times H$}{
              Observe $s_h^i = \{A_{\text{spec}}, D_{\text{region}}, G_{\text{avg}}\}$\;
              Take action $a_h^i = \pi_h(s_h^i)$ for $A_{\text{spec}}$\;
            }
          }

          \For{each region $j = 1$ to $B \times H \times R$}{
            Observe $s_l^j = \{A_{\text{spec}}, P_u, P_v, H, I\}$\;
            Take action $a_l^j = \pi_l(s_l^j)$ for $\beta_j$, $\alpha_j$, $\Delta p_j$\;
            Update UAV positions: $P_v = P_v + \Delta p$\;
          }

          \For{each region $j$}{
            Calculate local metrics\;
            \hspace{3em}$\{\gamma_{ju}$, $R_{ju}$, $\eta_j$, $F_j$, $V_j$\}\;
          }

          Compute rewards: $r_s$, $r_h$, $r_l$\;
          Store $(s_s, a_s, r_s)$ in $D_s$\;
          Store $(s_h, a_h, r_h)$ in $D_h$\;
          Store $(s_l, a_l, r_l)$ in $D_l$\;

          Update $\pi_s$, $\pi_h$, $\pi_l$ using samples from buffers\;

          \If{terminated or truncated}{
            \textbf{break}\;
          }
        }
        Calculate episodic metrics $R_{avg}$, $\eta$, $F$\;
      }
      \Return $\pi_s$, $\pi_h$, $\pi_l$, $R_{avg}$, $\eta$, $F$\;
    \end{algorithm}
  }





\section{Numerical Results}
\subsection{Experimental Settings}
The experimental framework utilized a hierarchical reinforcement learning environment with a detailed network topology. It featured $2$ beams from LEO, each with $1$ HAP, subdivided into $2$ regions, each with $2$ base stations and $1$ UAV, supporting $10$ users per region. Spectrum allocation spanned $200$ MHz across $10$ subbands centered at $28$ GHz. Node altitudes included satellites at $550$ km, HAPs at $20$ km, and UAVs at $100$ meters. Transmission power ranged from $33$–$45$ dBm for satellites, $28$–$36$ dBm for HAPs, $16$ dBm for base stations, and $8$ dBm for UAVs. Each $2 \times 2$ km region featured UAVs moving in $10$-meter steps, with noise power fixed at $-174$ dBm/Hz.

The reward mechanism used a weighted multi-objective structure: spectrum efficiency ($1.5$), fairness ($0.5$), UAV penalty ($-1.0$), and QoS violations ($-0.5$). Employed proximal policy optimization (PPO) as a single-agent RL algorithm with a learning rate of $0.0005$, mini-batch size of $512$, batch size of $2000$, $30$ stochastic gradient descent iterations, a discount factor of $0.99$, entropy coefficient of $0.01$, and value function loss coefficient of $1.0$. All experiments were run on i9-14900X 24 cores with 1x NVIDIA T4 16GB VRAM.

\subsection{Baseline Methodologies}

Results compare the proposed HDRL framework against SADRL \& MADRL algorithms used in spectrum sharing scenarios. A direct comparison with frameworks used in other relevant works, however, is unfeasible as the system model and components are vastly different. Evaluation considers a comprehensive SAG network as the default scenario, unless mentioned otherwise.

\begin{enumerate}
  \item \textbf{Exhaustive Search.} This baseline conducts a full combinatorial search across all possible spectrum allocation configurations to identify the optimal solution. However, the exhaustive search is computationally prohibitive for large-scale networks due to its exponential complexity.
  \item \textbf{Random Allocation.} Serving as a performance lower bound, this approach assigns spectrum resources randomly without leveraging any intelligent decision-making, offering a baseline for evaluating the gains achieved through reinforcement learning.
  \item \textbf{SADRL.} PPO is a SADRL algorithm that learns a global policy to optimize network objectives as is commonly used in literature.~\cite{10039119} relies on PPO to optimize spectrum sharing in an IRS-aided cognitive radio system, and~\cite{10667637} optimizes subcarrier spacing and uplink-downlink allocation with PPO in 5G networks.
  \item \textbf{MADRL.} Various works formulate MADRL algorithms to enhance spectrum sharing.~\cite{9426930} employs MADRL to enable cooperative spectrum sensing among multiple secondary users in cognitive radio networks, enhancing sensing accuracy by sharing detection results and reducing overhead through decentralized execution.~\cite{9329087} proposes a distributed resource management mechanism using MADRL and independent Q-learning to determine user scheduling and power allocation.
\end{enumerate}

\subsection{Results}
\begin{figure*}[t!]
  \centering
  \begin{subfigure}[t]{0.32\textwidth}
    \centering
    \includegraphics[width=\textwidth]{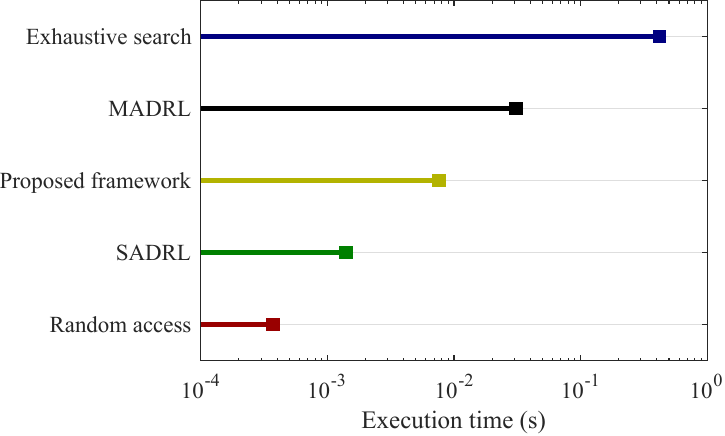}
    \caption{Execution time (s) for different algorithms.}
    \label{fig:subfig1}
  \end{subfigure}%
  \hspace{0.5em}
  \begin{subfigure}[t]{0.32\textwidth}
    \centering
    \includegraphics[width=\textwidth]{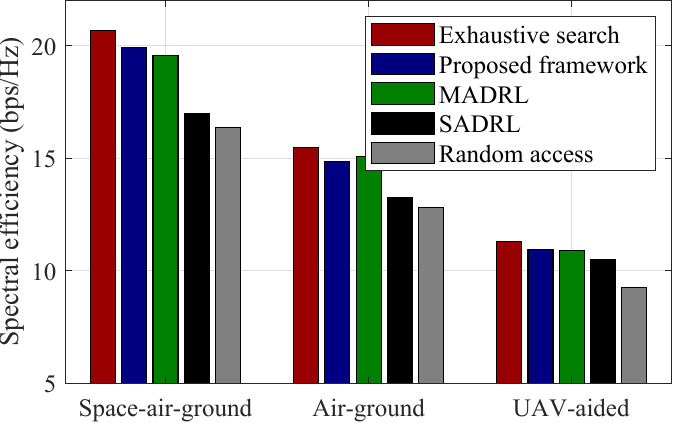}
    \caption{Spectral efficiency (bps/Hz) of algorithms in different network structures.}
    \label{fig:subfig2}
  \end{subfigure}%
  \hspace{0.5em}
  \begin{subfigure}[t]{0.32\textwidth}
    \centering
    \includegraphics[width=\textwidth]{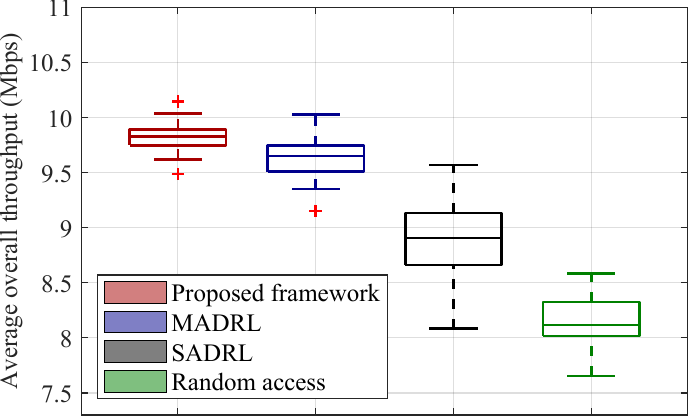}
    \caption{Average overall throughput (Mbps) against number of steps in evaluation}
    \label{fig:subfig3}
  \end{subfigure}


  \begin{subfigure}[t]{0.32\textwidth}
    \centering
    \includegraphics[width=\textwidth]{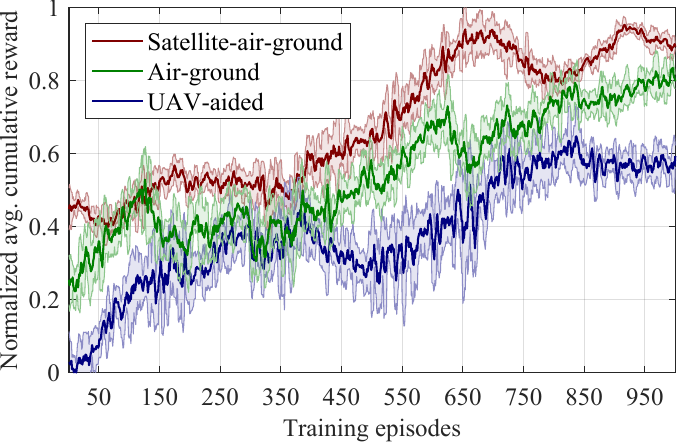}
    \caption{Normalized average cumulative in training for different network structures.}
    \label{fig:subfig4}
  \end{subfigure}%
  \hspace{0.5em}
  \begin{subfigure}[t]{0.32\textwidth}
    \centering
    \includegraphics[width=\textwidth]{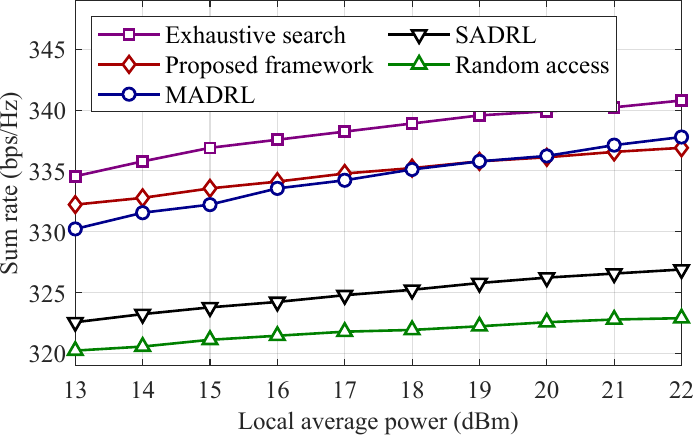}
    \caption{Sum rate (bps/Hz) of algorithms against local average power (dBm)}
    \label{fig:subfig5}
  \end{subfigure}%
  \hspace{0.5em}
  \begin{subfigure}[t]{0.32\textwidth}
    \centering
    \includegraphics[width=\textwidth]{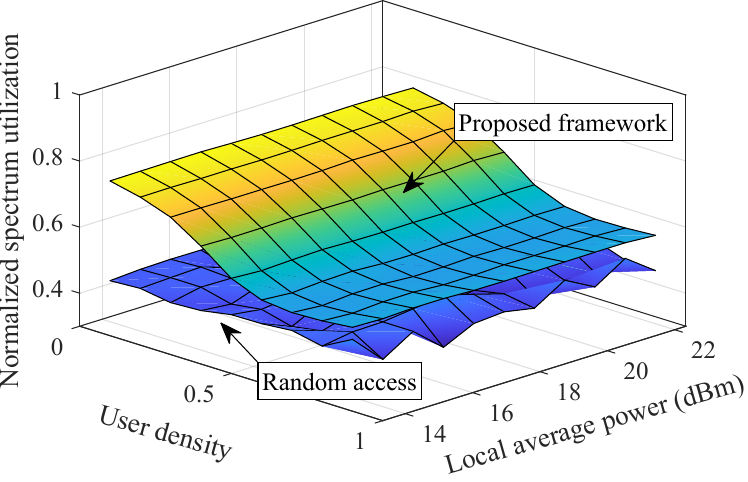}
    \caption{Normalized spectrum utilization of algorithms against normalized user density and local average power (dBm)}
    \label{fig:subfig6}
  \end{subfigure}

  \caption{Results compare the proposed framework with benchmark models for different metrics.}
  \label{fig:collection}
\end{figure*}

Fig.~\ref{fig:subfig1} compares execution times of spectrum allocation algorithms. Exhaustive search has the highest time due to evaluating all possible action-state combinations. MADRL is slower than HDRL due to the exponential growth of joint state-action spaces and non-stationary learning environments with multiple agents training simultaneously, compared to HDRL's structured task decomposition and transfer learning between hierarchical levels. The proposed HDRL framework achieves a balance, segmenting decision-making hierarchically to improve efficiency without compromising decision quality. The proposed framework is 3.75x faster than MADRL and 50x faster than exhaustive search. SADRL's centralized structure leads to faster execution, while random access, lacking intelligent decision-making, is the fastest. The HDRL framework is 4x and 19x slower than SADRL and random access, respectively.

Fig.~\ref{fig:subfig2} compares spectral efficiency across various network scenarios, illustrating the performance of different allocation algorithms. Exhaustive search achieves the highest efficiency possible in all network scenarios and the proposed framework achieves 97\% of that optimal spectral efficiency in SAG network, and achieves 2\% more than MADRL. This shows that the proposed framework achieves near-optimal performance while maintaining better computational efficiency. In the SAG network scenario, the proposed framework outperforms SADRL and random access by approximately 15\% and 18\%, respectively. The performance gap between all algorithms narrows in other scenarios due to reduced degrees of freedom and increased interference in simpler architectures. The difference is more pronounced in SAG networks because it has higher system complexity with three distinct layers offering more degrees of freedom to exploit.

Fig.~\ref{fig:subfig3} illustrates the throughput performance of different algorithms against varying channel conditions. The proposed framework maintains stable performance across all steps, averaging approximately 9.85 Mbps, while other approaches show more variance in their throughput values. MADRL closely follows with an average throughput of about 9.65 Mbps. The proposed framework achieves achieved 5\%, 11\% and 25\% higher average throughput than MADRL, SADRL, and random access, respectively. SADRL shows lower average throughput due to its limitations in capturing complex network dynamics and has the highest throughput fluctuations at every step. Random access performs worst with the lowest throughput, as expected from its non-intelligent randomized selection of actions.

Fig.~\ref{fig:subfig4} demonstrates the learning progression and convergence behavior of the proposed framework across different network architectures over 1000 training episodes. The SAG network achieves the highest normalized average cumulative reward due to its rich multi-layer structure offering more optimization opportunities. The air-ground (AG) network shows moderate performance, with a similar learning pattern but lower overall rewards due to reduced network complexity. The UAV-aided network, being the simplest architecture, exhibits the lowest reward values but shows steady improvement. All scenarios show initial fluctuations during training before stabilizing, with consistent performance gaps.

\begin{table}[t!]
  \centering
  \caption{Simulation parameters.}
  \label{tab:params}
  \resizebox{0.8\columnwidth}{!}{%
    \begin{tabular}{ll}
      \toprule
      \textbf{Parameter}                                      & \textbf{Value} \\
      \midrule
      Number of beams ($B$)                                   & 2              \\
      HAPs per beam ($H$)                                     & 1              \\
      Regions per HAP ($R$)                                   & 2              \\
      Time blocks per region ($T$)                            & 2              \\
      UAVs per region ($U$)                                   & 1              \\
      Users per region ($K$)                                  & 10             \\
      Total bandwidth ($W$)                                   & 200 MHz        \\
      Number of subbands ($N$)                                & 10             \\
      Maximum episodes ($M$)                                  & 1000           \\
      Steps per episode ($S$)                                 & 500            \\
      Decision intervals ($\delta_s$, $\delta_h$, $\delta_l$) & (50, 10, 1)    \\
      \bottomrule
    \end{tabular}%
  }
\end{table}

Fig.~\ref{fig:subfig5} illustrates the sum rate of various algorithms relative to the local average power, calculated as the average power across a region. The proposed framework and MADRL perform equally well. Other algorithms display consistent trends, with sum rates gradually increasing as average power rises. Exhaustive search consistently outperforms the proposed framework by 5 bps/Hz, equivalent to a 1.5\% improvement. Meanwhile, SADRL and random access remain close, within 4 bps/Hz of each other, but lag at least 8 bps/Hz behind the proposed framework.

Fig.~\ref{fig:subfig6} is a 3D surface plot illustrating the relationship between normalized spectrum utilization and, user density, and local average power. The color gradient represents utilization, with darker blue indicating lower and yellow-orange denoting higher utilization. The graph demonstrates the adaptability of the proposed framework, revealing that at user density close to $1$, spectrum utilization decreases. This reduction is attributed to the larger action space required to accommodate a greater number of users.


\section{Ablation Studies}

Extensive ablation studies provide deeper insights into HDRL's performance. These evaluate the impact of different reward function formulations, including cumulative reward to capture the total aggregated rewards over the episode trajectory, aggregative reward to normalize the cumulative reward by episode duration and to reduce the influence of varying episode lengths, and difference reward to isolate the unique contribution of individual agents to the overall system performance. Results use rigorous hyperparameter tuning to explore the most impactful parameters such as learning rate, PPO clip parameter, and entropy coefficient. Evaluation tests the robustness of the proposed HDRL solution by examining its performance under varying longitudinal and latitudinal conditions, ensuring its adaptability to diverse network configurations and environments. The ablation studies are extensively covered in Supplementary Materials at the end of this paper.

\section{Conclusion}

The growing demand for seamless wireless connectivity, further accelerated by 6G adoption, necessitates efficient spectrum allocation to serve diverse user demands. Existing spectrum-sharing solutions predominantly address TNs, overlooking the critical role of NTNs such as satellite constellations from SpaceX, OneWeb, and Amazon. The research introduces an HDRL framework for dynamic spectrum allocation within an integrated TN-NTN infrastructure. By leveraging network nesting, DRL agents at each tier of the network hierarchy coordinate spectrum management based on user demand. This adaptive, multi-layered approach surpasses algorithms like MADRL in efficiency and responsiveness, supporting the diverse connectivity requirements of 6G ecosystems and empowering network operators to manage spectrum resources.


\bibliography{main}

\end{document}